\documentclass[
    aps,
    prl,
    twocolumn,
    nofootinbib,
    preprintnumbers,
    superscriptaddress,
    altaffilletter
]{revtex4-2}

\usepackage{amsmath}
\usepackage{amsfonts}
\usepackage{amssymb}
\usepackage{mathrsfs}
\usepackage{mathdots}
\usepackage{mathtools}
\usepackage{mleftright}
\usepackage{graphicx}

\usepackage{xcolor}
\definecolor{LinkColor}{rgb}{0.06,
0.31,0.55}
\definecolor{CiteColor}{rgb}{0.0,0.55,0.0}
\definecolor{UrlColor}{rgb}{0.55,0.28,0.55}

\usepackage{hyperref}
\hypersetup{
    linktoc=all,
    colorlinks=true,
    linkcolor=LinkColor,
    citecolor=CiteColor,
    urlcolor=UrlColor
}

\newcommand{\LR}[1]{\left(#1\right)}
\newcommand{\LRa}[1]{\bigl(#1\bigr)}
\newcommand{\LRb}[1]{\Bigl(#1\Bigr)}

\newcommand{\dd}{\mathrm{d}}
\newcommand{\pp}{\partial}
\newcommand{\ii}{i}

\newcommand{\thalf}{\tfrac{1}{2}}
\newcommand{\eqq}{\equiv}
\newcommand{\peq}{\phantom{=}}
\newcommand{\nn}{\nonumber}
\newcommand{\abs}[1]{\left\lvert#1\right\vert}
\newcommand{\sign}{\operatorname{sign}}

\newcommand{\RR}{\mathbb{R}}

\newcommand{\CC}{\mathbb{C}}
\newcommand{\ZZ}{\mathbb{Z}}

\newcommand{\bulk}{\mathsf{B}}

\newcommand{\bulkConj}{\ddagger}
\newcommand{\bdryConj}{\dagger}

\newcommand{\genM}{M}
\newcommand{\genL}{L}
\newcommand{\genP}{P}

\newcommand{\diffL}{\mathcal{L}}
\newcommand{\diffP}{\mathcal{P}}

\newcommand{\op}{\mathcal{O}}
\newcommand{\ket}[1]{\lvert#1\rangle}

\newcommand{\vev}[1]{\langle#1\rangle}

\newcommand{\innout}{s}
\newcommand{\inn}{\texttt{-}}
\newcommand{\out}{\texttt{+}}
\newcommand{\typeT}{\texttt{T}}
\newcommand{\typeM}{\texttt{M}}

\newcommand{\mass}{m}
\newcommand{\branch}{\sigma}
\newcommand{\deltaC}{\delta_{\CC}}
\newcommand{\coThree}{C}
\newcommand{\cycsign}{\varsigma}

\begin{document}

\title{Carrollian holography with agentic AI: Real mass is imaginary}

\author{Reiko Liu}
\email{reiko.antoneva@foxmail.com}
\affiliation{Shanghai Institute for Mathematics and Interdisciplinary Sciences (SIMIS), Shanghai, 200433, China}

\author{Wen-Jie Ma}
\email{wenjie.ma@simis.cn}
\affiliation{Fudan Center for Mathematics and Interdisciplinary Study, Fudan University, Shanghai, 200433, China}
\affiliation{Shanghai Institute for Mathematics and Interdisciplinary Sciences (SIMIS), Shanghai, 200433, China}

\author{Hu Zheng}
\email{zheng.hu.263@gmail.com}
\noaffiliation

\author{Yu-fan Zheng}
\email{zhengyufan@bimsa.cn}
\affiliation{Beijing Institute of Mathematical Sciences and Applications (BIMSA), Huaibei Town, Huairou District, Beijing 101408, China}

\begin{abstract}
    We introduce LACIA, a verification-driven agentic AI workflow for theoretical physics, and apply it with independent human checks to construct Carrollian conformal bases.
    We develop the Poincare-Carrollian intertwiner as the central method. It reproduces the celestial and Carrollian conformal bases for massless particles and constructs the missing Carrollian bases for massive and tachyonic particles. The massive basis requires a complex momentum shift in scattering amplitudes.
\end{abstract}

\maketitle

\section{Introduction}

Carrollian and celestial holography seek to reorganize flat-space scattering amplitudes as correlators of boundary Carrollian and celestial CFTs \cite{Cheung:2016iub,Pasterski:2016qvg,Pasterski:2017kqt,Strominger:2017zoo,Raclariu:2021zjz,Pasterski:2021rjz,Pasterski:2021raf,McLoughlin:2022ljp} \cite{Banerjee:2018gce,Donnay:2022aba,Bagchi:2022emh,Nguyen:2025zhg}. In celestial holography, dictionaries for particles of arbitrary mass and spin have been constructed in \cite{Pasterski:2016qvg,Pasterski:2017kqt,Law:2020tsg,Iacobacci:2020por,Narayanan:2020amh,Chang:2022jut,Chang:2022seh,Chang:2023ttm,Liu:2025voe}. In Carrollian holography, the only known dictionary is the Mellin-Laplace transform for massless particles \cite{Banerjee:2018gce}, in which the extrapolation dictionary appears as a special case.

This leaves a long-standing open problem in Carrollian holography. Without a local-operator description of massive particles, the corresponding boundary correlators are not yet defined.

\begin{figure}[htbp!]
    \centering
    \includegraphics[width=\linewidth]{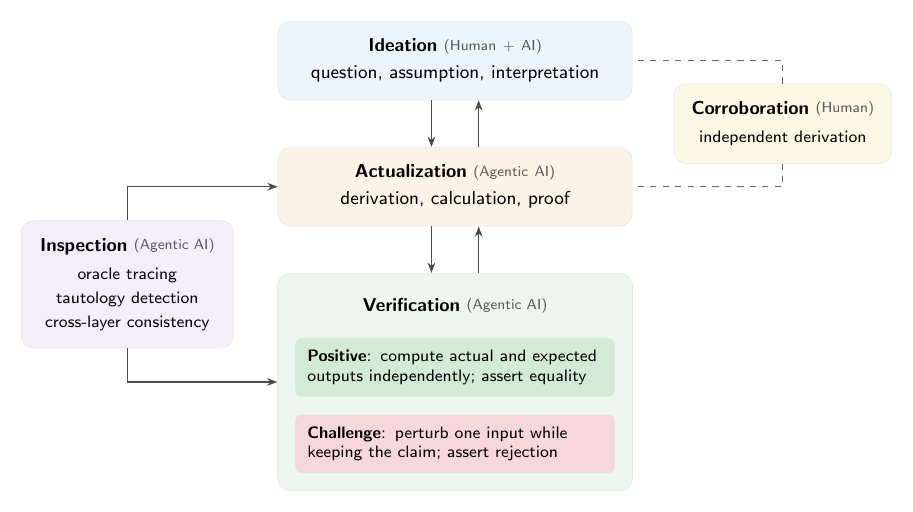}
    \caption{
        Schematic structure of the LACIA workflow.
        Corroboration is an external human-check layer.
    }
    \label{fig:lacia-workflow}
\end{figure}

We develop LACIA, an agentic AI workflow centered on the separation of physical reasoning from technical realization, designed to reduce the risk of hallucinated derivation and deceptive verification; see Figure~\ref{fig:lacia-workflow}.
We then apply LACIA to develop a systematic intertwiner method for constructing Carrollian conformal bases. As a starting point, we focus on three-dimensional scalar particles. The method reproduces the known massless bases and constructs the previously missing tachyonic and massive bases; the tachyonic basis has real momentum support, whereas the massive basis requires a complex shift of momentum.

\section{Agentic AI workflow: LACIA}

Theoretical physics proceeds through a loop of ideas and calculations. A physical idea is turned into a calculation; the result is then used to revise the idea. When this loop is carried by a single researcher, its internal boundary can remain implicit. In a human-AI system, the boundary becomes an interface. This motivates us to develop the LACIA workflow, a Language-model Agent Cycle for Ideation and Actualization.

In LACIA, ideation is separated from actualization. Ideation is the layer of physical reasoning: it fixes the question, assumption, and interpretation. Actualization is the layer of technical realization: it carries out analytic derivation and invokes symbolic or formal tools for calculation and proof.

The separation is not enough. Actualization is carried out by agentic AI, a probabilistic language-model system, and long technical derivations can easily contain hallucinated steps \cite{kalai2025}. To reduce this risk, actualization must be constrained by symbolic or formal verification. Verification turns technical claims into fail-capable checks, thereby improving the reliability of the result.

Nor is verification enough. Verification is also carried out by agentic AI and can be affected by deceptive behavior such as reward hacking \cite{baker2025}, producing, for example, tautological verification rather than genuine evidence. This requires inspection, which checks whether verification has actually tested the actualization.

This leads to four minimal layers of LACIA: ideation formulates physical reasoning, actualization performs technical realization, verification tests actualized claims, and inspection examines verification itself. \emph{Human-AI trust becomes possible only through structured mutual distrust.}

In the present work, LACIA is implemented at the skill level in Codex (GPT-5.5) and used throughout the physical calculations. The workflow produces about 40 verification units and 20k lines of SymPy code, which will be made publicly available later. The key physical computations are also performed independently by the human authors. Hence the verification artifacts and the human calculations provide two independent lines of confirmation.

\section{Beyond change of basis}

Carrollian and celestial holography are not merely changes of basis.
The reason is structural: although the bulk and boundary symmetries are matched, their natural time evolutions are different, leading to inequivalent quantizations and Hilbert spaces.

In the bulk, states belong to unitary representations of the isometry algebra, and their conjugation comes from bulk time evolution.
On the boundary, local operators and their derivatives lie in representations of the conformal algebra, and are equipped with the BPZ conjugation from the boundary radial quantization.
These two types of representations are not isomorphic, even though they may share the same infinitesimal transformations \cite{Chen:2022cpx}; for more mathematical discussions, see \cite[Appendix A]{cap2000} and \cite[Section 2]{kobayashi2015}.

More precisely, in celestial holography, a single-particle state is decomposed into a direct integral of principal series representations of the Lorentz algebra, then each principal series is mapped to a Verma module through an intertwiner. In this way, a single-particle state is equivalently encoded by primary operators with tunable conformal dimension. This is the representation-theoretic content behind the familiar celestial basis.

Similarly, a Carrollian basis must be an intertwiner from a Poincare representation to a Carrollian conformal module. The existence of such an intertwiner is already implicit in the Carrollian shadow formalism \cite{Chen:2022cpx}.

\paragraph{Poincare and Carrollian conformal symmetries.}

Take the bulk metric to be $g=\operatorname{diag}(-,+,+)$. The three-dimensional Poincare algebra is
\begin{align}
    \nn
    [\genM_{ab},\genM_{cd}]
    &=\genM_{ad}g_{bc}+\genM_{bc}g_{ad}
    -\genM_{ac}g_{bd}-\genM_{bd}g_{ac}
    ,
    \\
    \label{eq:poincare-algebra}
    [\genM_{ab},\genP_c]
    &=\genP_a g_{bc}-\genP_b g_{ac}
    ,
\end{align}
whereas the boundary Carrollian conformal algebra is
\begin{equation}
    \label{eq:carrollian-algebra}
    \begin{aligned}
    [\genL_n,\genL_m]&=(n-m)\genL_{n+m}
    ,
    \\
    [\genL_n,\genP_m]&=(n-m)\genP_{n+m}
    ,
    \end{aligned}
\end{equation}
for $m,n=-1,0,1$.
At null infinity, the bulk isometry must be matched to the boundary Carrollian conformal symmetry. Computing the (conformal) Killing vector fields on the two sides gives the correct isomorphism, matching with \cite{Chen:2022cpx,Chen:2023naw}:
\begin{equation}
    \label{eq:poincare-carrollian-map}
    \begin{aligned}
    \genL_- &= \genM_{01}+\genM_{12},
    &
    \genP_- &= \genP_0-\genP_2
    ,
    \\
    \genL_0 &= -\genM_{02},
    &
    \genP_0 &= -\genP_1
    ,
    \\
    \genL_+ &= \genM_{12}-\genM_{01},
    &
    \genP_+ &= \genP_0+\genP_2
    .
    \end{aligned}
\end{equation}
Here we have slightly abused notation: $\genL_{\pm}\eqq \genL_{\pm 1}$ and left/right sides are boundary/bulk generators.

\paragraph{Conjugation.}

The difference between the bulk and boundary state-space structures is already visible at the level of conjugation.
Using \eqref{eq:poincare-carrollian-map}, the conjugation inherited from bulk unitarity is
\begin{equation}
    \label{eq:bulk-conjugation}
    \genL_n^{\bulkConj}=-\genL_n
    ,
    \quad
    \genP_n^{\bulkConj}=-\genP_n
    ,
\end{equation}
whereas the Carrollian Verma module carries the BPZ conjugation
\begin{equation}
    \label{eq:boundary-conjugation}
    \genL_n^{\bdryConj}=\genL_{-n}
    ,
    \quad
    \genP_n^{\bdryConj}=\genP_{-n}
    .
\end{equation}

Thus the two sides have compatible symmetry actions but different adjoint structures. This is one consequence of the statement that the Carrollian dictionary is not merely a change of basis.

\section{Intertwiner method}

An intertwiner is a linear map between different representations of isomorphic algebras that preserves the transformation laws.

\paragraph{Bulk scalar representation.}

The source of the intertwiner is a scalar representation of the Poincare algebra,
\begin{equation}
    \label{eq:Poincare-algebra-scalar-module}
    [\genM_{ab}, \op^{\bulk}]
    =
    \LRa{
        q_{b} \pp_{a} - q_{a} \pp_{b}
    }
    \op^{\bulk}
    ,
    \quad
    [\genP_{a}, \op^{\bulk}]
    =
    \ii \innout q_{a} \op^{\bulk}
    ,
\end{equation}
where $\op^{\bulk}$ denotes the creation/annihilation operator, $q$ is the momentum, and the incoming/outgoing sign is
\begin{equation}
    \label{eq:inout-sign}
    \innout
    =
    \begin{dcases}
        +1, & \text{outgoing}, \\
        -1, & \text{incoming}.
    \end{dcases}
\end{equation}
Using the isomorphism \eqref{eq:poincare-carrollian-map}, the Poincare action can be rewritten as
\begin{equation}
    \label{eq:bulk-action}
    [\genL_n,\op^{\bulk}]
    =\diffL_n^{\bulk}\op^{\bulk}
    ,
    \quad
    [\genP_n,\op^{\bulk}]
    =\diffP_n^{\bulk}\op^{\bulk}
    .
\end{equation}

For a massless particle, the momentum is
\begin{equation}
    \label{eq:massless-shell}
    q
    =
    \omega
    (1+z^2,2z,1-z^2)
    ,
\end{equation}
with energy $\omega>0$ and angular variable $z\in\RR$. The bulk differential operators in \eqref{eq:bulk-action} are
\begin{equation}
    \label{eq:massless-bulk-action}
    \diffL_n^{\bulk}
    =
    z^{n+1}\pp_{z}
    -(n+1)z^n\omega \pp_{\omega}
    ,
    \quad
    \diffP_n^{\bulk}
    =
    -2\ii\innout \omega z^{n+1}
    .
\end{equation}

For a massive particle, the momentum is
\begin{equation}
    \label{eq:massive-shell}
    q
    =
    \tfrac{\mass}{2y}
    (y^2+z^2+1,2z,-y^2-z^2+1)
    ,
\end{equation}
with mass $\mass>0$ and energy variable $y>0$. The bulk differential operators in \eqref{eq:bulk-action} are
\begin{align}
    \diffL_n^{\bulk}
    &=
    z^{n+1}\pp_{z}
    +(n+1)z^n y\pp_{y}
    -\thalf n(n+1)z^{n-1}y^2\pp_{z}
    ,
    \nn
    \\
    \label{eq:massive-bulk-action}
    \diffP_n^{\bulk}
    &=
    -\ii\innout\mass y^{-1}
    \LRa{z^{n+1}+\thalf n(n+1)z^{n-1}y^2}
    .
\end{align}

The tachyonic momentum is
\begin{equation}
    \label{eq:tachyonic-shell}
    q
    =
    \tfrac{\mass}{2y}
    (-y^2+z^2+1,2z,y^2-z^2+1)
    ,
\end{equation}
with mass parameter $\mass>0$. Here ``tachyonic particle'' means an excitation in a tachyonic unitary Poincare representation, rather than a canonically quantized tachyonic scalar field. It has no incoming/outgoing distinction, and accordingly the energy variable $y$ is sign-indefinite, unlike in the massive case. The bulk differential operators in \eqref{eq:bulk-action} are
\begin{align}
    \diffL_n^{\bulk}
    &=
    z^{n+1}\pp_{z}
    +(n+1)z^n y\pp_{y}
    +\thalf n(n+1)z^{n-1}y^2\pp_{z}
    ,
    \nn
    \\
    \label{eq:tachyonic-bulk-action}
    \diffP_n^{\bulk}
    &=
    -\ii\mass y^{-1}
    \LRa{z^{n+1}-\thalf n(n+1)z^{n-1}y^2}
    .
\end{align}

\paragraph{Carrollian Verma module.}

The target of the intertwiner is a primary operator $\op_{h,\xi}(z,u)$ with coordinates $(z,u)$ and weights $(h,\xi)$ in Carrollian CFT. Lying in a Verma module, the corresponding primary state $\ket{h,\xi}$ is characterized by
\begin{equation}
    \label{eq:carrollian-primary-state}
    \begin{aligned}
    \genL_0\ket{h,\xi}
    &=
    h\ket{h,\xi},
    &
    \genP_0\ket{h,\xi}
    &=
    \xi\ket{h,\xi}
    ,
    \\
    \genL_+\ket{h,\xi}
    &=
    0,
    &
    \genP_+\ket{h,\xi}
    &=
    0
    .
    \end{aligned}
\end{equation}
The infinitesimal transformations of $\op_{h,\xi}$ are
\begin{equation}
    \label{eq:boundary-action}
    [\genL_n,\op_{h,\xi}]=\diffL_n\op_{h,\xi}
    ,
    \quad
    [\genP_n,\op_{h,\xi}]=\diffP_n\op_{h,\xi}
    ,
\end{equation}
where the boundary differential operators are
\begin{equation}
    \label{eq:boundary-differentials}
    \begin{aligned}
    \diffL_n
    &=
    z^{n+1}\pp_{z}
    +(n+1)u z^n\pp_{u}
    \\
    &\peq
    +(n+1)h z^n
    +n(n+1)\xi u z^{n-1}
    ,
    \\
    \diffP_n
    &=
    z^{n+1}\pp_{u}
    +(n+1)\xi z^n
    .
    \end{aligned}
\end{equation}

\paragraph{Intertwiner.}

After preparing the bulk and boundary representations, we define the Carrollian basis as the integral transform
\begin{equation}
    \label{eq:kernel-ansatz}
    \op_{h,\xi}(z,u)
    =
    \int \dd\mu(\lambda')
    K_{h,\xi}(z,u;\lambda')
    \op^{\bulk}(\lambda')
    ,
\end{equation}
where the kernel $K_{h,\xi}$ is to be determined.
Here $\lambda'$ denotes the momentum variables being integrated over, and to simplify the intertwiner equations below, we choose the measure to be Lorentz-invariant,
\begin{equation}
    \label{eq:invariant-measures}
    \begin{alignedat}{3}
        \text{massless}
        &\quad&
        \lambda'
        &=(\omega',z')
        \quad&
        \dd\mu
        &=\dd\omega' \dd z'
        ,
        \\
        \text{massive}
        &\quad&
        \lambda'
        &=(y',z')
        \quad&
        \dd\mu
        &=y'^{-2}\dd y' \dd z'
        ,
        \\
        \text{tachyonic}
        &\quad&
        \lambda'
        &=(y',z')
        \quad&
        \dd\mu
        &=y'^{-2}\dd y' \dd z'
        .
    \end{alignedat}
\end{equation}
We then match the transformations \eqref{eq:bulk-action} and \eqref{eq:boundary-action} on both sides of \eqref{eq:kernel-ansatz}, and perform integration by parts to obtain the intertwiner equations
\begin{align}
    \label{eq:intertwiner-equation-Lorentz}
    \LRa{\diffL_n+\diffL_n^{\prime\bulk}}
    K_{h,\xi}&=0
    ,
    \\
    \label{eq:intertwiner-equation-translation}
    \LRa{\diffP_n-\diffP_n^{\prime\bulk}}
    K_{h,\xi}&=0
    .
\end{align}
The primes indicate that bulk differential operators act on primed variables $\lambda'$.

\paragraph{Massless benchmark.}

To validate the intertwiner method, we reproduce the known massless basis. If omitting the translation equation \eqref{eq:intertwiner-equation-translation} and only imposing the Lorentz equation \eqref{eq:intertwiner-equation-Lorentz}, the solutions are the celestial Mellin basis
\begin{equation}
    \label{eq:mellin-kernel}
    K_{h}
    =
    \omega'^{h-1}\delta(z'-z)
    ,
\end{equation}
and the shadow basis
\begin{equation}
    \label{eq:shadow-kernel}
    K_{h}
    =
    \omega'^{-h}
    \abs{z'-z}^{-2h}
    .
\end{equation}
Translations shift the conformal weight $h$, and hence are not endomorphisms of a fixed Verma module \cite{Law:2019glh,Pate:2019lpp}.

Imposing both equations \eqref{eq:intertwiner-equation-Lorentz} and \eqref{eq:intertwiner-equation-translation}, there are no regular solutions and two types of distributional solutions.
The first one exists only for $\xi=0$ and gives the Mellin-Laplace basis,
\begin{equation}
    \label{eq:mellin-laplace-kernel}
    K^{\innout}_{h,0}
    =
    \omega'^{h-1}
    \exp(-2\ii\innout\omega'u)
    \delta(z'-z)
    .
\end{equation}
Thus a massless particle is mapped to a primary operator with unconstrained conformal weights $(h,0)$. At nonpositive integer $h$, the Mellin factor $\omega'^{h-1}$ can descend to the delta function $\delta^{(-h)}(\omega')$, giving a discrete soft basis similar to \cite{Mitra:2024ugt}.

The other ``soft shadow'' solution exists only for $h=1$ and $\xi=0$,
\begin{equation}
    \label{eq:soft-shadow-kernel}
    K^{\text{shadow}}_{1,0}
    =
    \delta(\omega')
    (z'-z)^{-2}
    ,
\end{equation}
which is related to the $h=0$ soft basis by a Carrollian shadow transform.

\section{Tachyonic and massive bases}

Although tachyonic particles are not physical asymptotic states in the usual sense, they are indispensable ingredients in Carrollian shadow formalism \cite{Chen:2022cpx}, celestial split representation \cite{Chang:2023ttm,Liu:2024lbs,Liu:2024vmx} and regular celestial amplitude \cite{Liu:2025dhh,Liu:2025voe,Liu:2026ocv}. We therefore begin by constructing the tachyonic Carrollian basis.

\paragraph{Tachyonic particle.}

The $\genL_-$ and $\genP_-$ equations in \eqref{eq:intertwiner-equation-Lorentz} and \eqref{eq:intertwiner-equation-translation} fix the $u$-dependence,
\begin{equation}
    K^{\typeT}
    \sim
    f(y',z'-z)
    \exp\LR{-\ii\mass y'^{-1}u}
    ,
\end{equation}
then the $\genL_0$ and $\genP_0$ equations exclude regular solutions and force the distributional form
\begin{equation}
    \label{eq:tachyonic-prekernel}
    K^{\typeT}
    \sim
    y'^{1-h}
    \delta\LRb{
        z'-z+\frac{\xi y'}{\ii\mass}
    }
    \exp(-\ii\mass y'^{-1}u)
    .
\end{equation}
The remaining $\genL_+$ and $\genP_+$ equations constrain the conformal weights to be
\begin{equation}
    \label{eq:tachyonic-weights}
    h=1
    ,
    \quad
    \xi=\branch \ii\mass
    ,
\end{equation}
where $\branch=\pm1$. This condition is precisely the matching of Casimir eigenvalues of the bulk and boundary representations.
Therefore the Carrollian basis is
\begin{equation}
    \label{eq:tachyonic-kernel}
    \begin{aligned}
    K^{\typeT}_{1,\branch\ii\mass}
    &=
    \tfrac{\mass}{2}
    \delta(z'-z+\branch y')
    \exp(-\ii\mass y'^{-1}u)
    .
    \end{aligned}
\end{equation}
The two branches $\branch=\pm1$ of the basis are related by the Carrollian shadow transform.

\paragraph{Massive particle.}
Similarly, the first four equations give
\begin{equation}
    \label{eq:massive-prekernel}
    K^{\typeM,\innout}
    \sim
    y'^{1-h}
    \delta\LRb{
        z'-z+\frac{\xi y'}{\ii\innout\mass}
    }
    \exp(-\ii\innout\mass y'^{-1}u)
    ,
\end{equation}
and the remaining two equations require
\begin{equation}
    \label{eq:massive-weights}
    h=1
    ,
    \quad
    \xi=\branch\mass
    ,
\end{equation}
where $\branch=\pm1$.
Therefore the Carrollian basis is
\begin{equation}
    \label{eq:massive-kernel}
    \begin{aligned}
    K^{\typeM,\innout}_{1,\branch\mass}
    &=
    \tfrac{\mass}{2}
    \deltaC(z'-z-\branch\innout\ii y')
    \exp(-\ii\innout\mass y'^{-1}u)
    .
    \end{aligned}
\end{equation}
Here $\deltaC$ is the complex-support delta function, an analytic functional \cite{Gelfand1,Gelfand2,Morimoto1} that has been used in celestial holography \cite{Donnay:2020guq,Liu:2024lbs,Liu:2025dhh,Liu:2025voe}. Equivalently, in the integral transform \eqref{eq:kernel-ansatz} it acts as a complex shift of the integrated momentum.

\paragraph{Massless limit.}

The normalizations in \eqref{eq:tachyonic-kernel} and \eqref{eq:massive-kernel} are chosen to simplify the massless limit.
Similar to the celestial case \cite{Liu:2025voe}, we perform the change of variable $y'=\frac{\mass}{2\omega'}$ for the massive basis and then take the massless limit, leading to
\begin{equation}
    \label{eq:massive-massless-limit}
    \lim_{\mass\to0}
    K^{\typeM,\innout}_{1,\branch\mass} \dd\mu
    =
    K^{\innout}_{1,0} \dd\mu
    .
\end{equation}
For the tachyonic basis, we split the energy into $y'>0$ and $y'<0$ and then use $y'=\pm\frac{\mass}{2\omega'}$, giving a linear combination of incoming and outgoing massless bases,
\begin{equation}
    \label{eq:tachyonic-massless-limit}
    \begin{aligned}
    \lim_{\mass\to0}
    K^{\typeT}_{1,\ii\mass} \dd\mu
    =
    (K^{\out}_{1,0}+K^{\inn}_{1,0}) \dd\mu
    .
    \end{aligned}
\end{equation}

\paragraph{Reality property.}

Comparing \eqref{eq:tachyonic-kernel} with \eqref{eq:massive-kernel}, the Poincare-Carrollian intertwiner exchanges the reality properties of mass and momentum. A tachyonic particle has imaginary mass and gives a Carrollian basis with real momentum support, whereas a massive particle has real mass and requires a complex momentum shift. This is the precise sense in which \emph{real mass is imaginary}.

\section{Future directions}

\paragraph{Carrollian amplitude.}

Having constructed the Carrollian basis for massive scalar particles, we can now define Carrollian amplitudes as
\begin{equation}
    \label{eq:carrollian-amplitude-definition}
    \vev{
        \op_{1}
        \cdots
        \op_{n}
    }
    =
    \prod_{i=1}^{n}
    \int K_{i} \dd\mu_i\,
    \mathcal{T}_n
    ,
\end{equation}
where $\mathcal{T}_n$ is the bulk scattering amplitude and $K_i$ denotes the $i$-th Carrollian basis.

For the process with one tachyonic and two massless particles, $\mathcal{T}_3=\delta^{(3)}(q_{1}{-}q_{2}{+}q_{3})$, the Carrollian amplitude takes the standard three-point form. The corresponding coefficient reads%
\footnote{
    Due to the topology of the compactified spatial coordinate $z\in \mathbb{RP}^{1}$, $2d$ three-point Carrollian correlators can depend on the cyclic order of operator insertions $\cycsign = \sign(z_{1,2} z_{2,3} z_{1,3})$.
    From the bulk perspective, $3d$ scattering amplitudes can depend on an extra $\ZZ_{2}$ quantum number $\sign(\epsilon_{abc} q_1^{a} q_2^{b} q_3^{c})$ \cite{Jain:2014nza,Inbasekar:2015tsa}, which is exactly $\cycsign$ here.
}
\begin{equation}
    \coThree^{\cycsign=+1}(
        \op^{\typeT}_{1,\ii\mass}
        \op^{\inn}_{h_2,0}
        \op^{\out}_{h_3,0}
    )
    =
    2^{-h_{2}-h_{3}}\mass^{h_{2}+h_{3}-2}
    ,
\end{equation}
and $\coThree^{\cycsign=-1} = 0$.
This provides a nontrivial check that the constructed basis transforms covariantly.

For massive particles, due to the complex momentum shift, the definition of Carrollian amplitudes \eqref{eq:carrollian-amplitude-definition} needs further refinement. We will give a more complete prescription in a forthcoming paper, and discuss the Carrollian OPE and conformal-block expansion using the techniques developed in \cite{Chen:2020vvn,Chen:2022cpx,Chen:2022jhx}.

\paragraph{4d and spin.}

Another natural next step is to construct Carrollian bases for four-dimensional and spinning particles. The key question is then to construct the target module on the Carrollian side. Candidates can be found in \cite{Chen:2021xkw,Chen:2023pqf,Chen:2024voz}, and especially in \cite{Chen:2023naw}.

\paragraph{Workflow benchmark.}

A separate future direction is to benchmark the LACIA workflow. Controlled experiments should compare this workflow with naive vibe research on theoretical-physics tasks, measuring the improvement in correctness and the reduction in hallucinated derivation and deceptive verification.

\vspace{\baselineskip}
\begin{acknowledgments}
    WJM is supported by the National Natural Science Foundation of China No. 12405082 and Shanghai Pujiang Program No. 24PJA118.
\end{acknowledgments}


\bibliography{ref}

@article{Jain:2014nza,
    title         = {{Unitarity, Crossing Symmetry and Duality of the S-matrix in large N Chern-Simons theories with fundamental matter}},
    author        = {Jain, Sachin and Mandlik, Mangesh and Minwalla, Shiraz and Takimi, Tomohisa and Wadia, Spenta R. and Yokoyama, Shuichi},
    eprint        = {1404.6373},
    year          = {2015},
    archiveprefix = {arXiv},
    doi           = {10.1007/JHEP04(2015)129},
    journal       = {JHEP},
    pages         = {129},
    primaryclass  = {hep-th},
    reportnumber  = {TIFR-TH-14-12, HRI-ST-1405, ICTS-2014-04},
    volume        = {04},
}

@article{Inbasekar:2015tsa,
    title         = {{Unitarity, crossing symmetry and duality in the scattering of $ \mathcal{N}=1 $ susy matter Chern-Simons theories}},
    author        = {Inbasekar, Karthik and Jain, Sachin and Mazumdar, Subhajit and Minwalla, Shiraz and Umesh, V. and Yokoyama, Shuichi},
    eprint        = {1505.06571},
    year          = {2015},
    archiveprefix = {arXiv},
    doi           = {10.1007/JHEP10(2015)176},
    journal       = {JHEP},
    pages         = {176},
    primaryclass  = {hep-th},
    reportnumber  = {TIFR-TH-15-15},
    volume        = {10},
}

@article{Cheung:2016iub,
    title         = {{4D scattering amplitudes and asymptotic symmetries from 2D CFT}},
    author        = {Cheung, Clifford and de la Fuente, Anton and Sundrum, Raman},
    eprint        = {1609.00732},
    year          = {2017},
    archiveprefix = {arXiv},
    doi           = {10.1007/JHEP01(2017)112},
    journal       = {JHEP},
    pages         = {112},
    primaryclass  = {hep-th},
    reportnumber  = {CALT-TH-2016-024, UMD-PP-017-010},
    volume        = {01},
}

@article{Pasterski:2016qvg,
    title         = {{Flat Space Amplitudes and Conformal Symmetry of the Celestial Sphere}},
    author        = {Pasterski, Sabrina and Shao, Shu-Heng and Strominger, Andrew},
    eprint        = {1701.00049},
    year          = {2017},
    archiveprefix = {arXiv},
    doi           = {10.1103/PhysRevD.96.065026},
    journal       = {Phys. Rev. D},
    number        = {6},
    pages         = {065026},
    primaryclass  = {hep-th},
    volume        = {96},
}

@article{Pasterski:2017kqt,
    title         = {{Conformal basis for flat space amplitudes}},
    author        = {Pasterski, Sabrina and Shao, Shu-Heng},
    eprint        = {1705.01027},
    year          = {2017},
    archiveprefix = {arXiv},
    doi           = {10.1103/PhysRevD.96.065022},
    journal       = {Phys. Rev. D},
    number        = {6},
    pages         = {065022},
    primaryclass  = {hep-th},
    volume        = {96},
}

@article{Banerjee:2018gce,
    title         = {{Null Infinity and Unitary Representation of The Poincare Group}},
    author        = {Banerjee, Shamik},
    eprint        = {1801.10171},
    year          = {2019},
    archiveprefix = {arXiv},
    doi           = {10.1007/JHEP01(2019)205},
    journal       = {JHEP},
    pages         = {205},
    primaryclass  = {hep-th},
    volume        = {01},
}

@article{Law:2019glh,
    title         = {{Poincare constraints on celestial amplitudes}},
    author        = {Law, Y. T. Albert and Zlotnikov, Michael},
    eprint        = {1910.04356},
    year          = {2020},
    archiveprefix = {arXiv},
    doi           = {10.1007/JHEP03(2020)085},
    journal       = {JHEP},
    note          = {[Erratum: JHEP 04, 202 (2020)]},
    pages         = {085},
    primaryclass  = {hep-th},
    volume        = {03},
}

@article{Pate:2019lpp,
    title         = {{Celestial operator products of gluons and gravitons}},
    author        = {Pate, Monica and Raclariu, Ana-Maria and Strominger, Andrew and Yuan, Ellis Ye},
    eprint        = {1910.07424},
    year          = {2021},
    archiveprefix = {arXiv},
    doi           = {10.1142/S0129055X21400031},
    journal       = {Rev. Math. Phys.},
    number        = {09},
    pages         = {2140003},
    primaryclass  = {hep-th},
    volume        = {33},
}

@article{Law:2020tsg,
    title         = {{Massive Spinning Bosons on the Celestial Sphere}},
    author        = {Law, Y. T. Albert and Zlotnikov, Michael},
    eprint        = {2004.04309},
    year          = {2020},
    archiveprefix = {arXiv},
    doi           = {10.1007/JHEP06(2020)079},
    journal       = {JHEP},
    pages         = {079},
    primaryclass  = {hep-th},
    volume        = {06},
}

@article{Donnay:2020guq,
    title         = {{Asymptotic Symmetries and Celestial CFT}},
    author        = {Donnay, Laura and Pasterski, Sabrina and Puhm, Andrea},
    eprint        = {2005.08990},
    year          = {2020},
    archiveprefix = {arXiv},
    doi           = {10.1007/JHEP09(2020)176},
    journal       = {JHEP},
    pages         = {176},
    primaryclass  = {hep-th},
    reportnumber  = {CPHT-RR022.042020},
    volume        = {09},
}

@article{Iacobacci:2020por,
    title         = {{Conformal Primary Basis for Dirac Spinors}},
    author        = {Iacobacci, Lorenzo and M{\"u}ck, Wolfgang},
    eprint        = {2009.02938},
    year          = {2020},
    archiveprefix = {arXiv},
    doi           = {10.1103/PhysRevD.102.106025},
    journal       = {Phys. Rev. D},
    number        = {10},
    pages         = {106025},
    primaryclass  = {hep-th},
    volume        = {102},
}

@article{Narayanan:2020amh,
    title         = {{Massive Celestial Fermions}},
    author        = {Narayanan, Sruthi A.},
    eprint        = {2009.03883},
    year          = {2020},
    archiveprefix = {arXiv},
    doi           = {10.1007/JHEP12(2020)074},
    journal       = {JHEP},
    pages         = {074},
    primaryclass  = {hep-th},
    volume        = {12},
}

@article{Chen:2020vvn,
    title         = {{On Galilean conformal bootstrap}},
    author        = {Chen, Bin and Hao, Peng-Xiang and Liu, Reiko and Yu, Zhe-Fei},
    eprint        = {2011.11092},
    year          = {2021},
    archiveprefix = {arXiv},
    doi           = {10.1007/JHEP06(2021)112},
    journal       = {JHEP},
    pages         = {112},
    primaryclass  = {hep-th},
    volume        = {06},
}

@article{Raclariu:2021zjz,
    title         = {{Lectures on Celestial Holography}},
    author        = {Raclariu, Ana-Maria},
    eprint        = {2107.02075},
    year          = {2021},
    archiveprefix = {arXiv},
    journal       = {},
    month         = {7},
    primaryclass  = {hep-th},
}

@article{Pasterski:2021rjz,
    title         = {{Lectures on celestial amplitudes}},
    author        = {Pasterski, Sabrina},
    eprint        = {2108.04801},
    year          = {2021},
    archiveprefix = {arXiv},
    doi           = {10.1140/epjc/s10052-021-09846-7},
    journal       = {Eur. Phys. J. C},
    number        = {12},
    pages         = {1062},
    primaryclass  = {hep-th},
    volume        = {81},
}

@article{Chen:2021xkw,
    title         = {{On higher-dimensional Carrollian and Galilean conformal field theories}},
    author        = {Chen, Bin and Liu, Reiko and Zheng, Yu-fan},
    eprint        = {2112.10514},
    year          = {2023},
    archiveprefix = {arXiv},
    doi           = {10.21468/SciPostPhys.14.5.088},
    journal       = {SciPost Phys.},
    number        = {5},
    pages         = {088},
    primaryclass  = {hep-th},
    volume        = {14},
}

@article{Donnay:2022aba,
    title         = {{Carrollian Perspective on Celestial Holography}},
    author        = {Donnay, Laura and Fiorucci, Adrien and Herfray, Yannick and Ruzziconi, Romain},
    eprint        = {2202.04702},
    year          = {2022},
    archiveprefix = {arXiv},
    doi           = {10.1103/PhysRevLett.129.071602},
    journal       = {Phys. Rev. Lett.},
    number        = {7},
    pages         = {071602},
    primaryclass  = {hep-th},
    volume        = {129},
}

@article{Bagchi:2022emh,
    title         = {{Scattering Amplitudes: Celestial and Carrollian}},
    author        = {Bagchi, Arjun and Banerjee, Shamik and Basu, Rudranil and Dutta, Sudipta},
    eprint        = {2202.08438},
    year          = {2022},
    archiveprefix = {arXiv},
    doi           = {10.1103/PhysRevLett.128.241601},
    journal       = {Phys. Rev. Lett.},
    number        = {24},
    pages         = {241601},
    primaryclass  = {hep-th},
    volume        = {128},
}

@article{Chen:2022cpx,
    title         = {{The shadow formalism of Galilean CFT$_{2}$}},
    author        = {Chen, Bin and Liu, Reiko},
    eprint        = {2203.10490},
    year          = {2023},
    archiveprefix = {arXiv},
    doi           = {10.1007/JHEP05(2023)224},
    journal       = {JHEP},
    pages         = {224},
    primaryclass  = {hep-th},
    volume        = {05},
}

@article{McLoughlin:2022ljp,
    title         = {{The SAGEX review on scattering amplitudes chapter 11: soft theorems and celestial amplitudes}},
    author        = {McLoughlin, Tristan and Puhm, Andrea and Raclariu, Ana-Maria},
    eprint        = {2203.13022},
    year          = {2022},
    archiveprefix = {arXiv},
    doi           = {10.1088/1751-8121/ac9a40},
    journal       = {J. Phys. A},
    number        = {44},
    pages         = {443012},
    primaryclass  = {hep-th},
    reportnumber  = {SAGEX-22-12, CPHT-RR016.032022, HU-EP-22/13, TCDMATH 22-02},
    volume        = {55},
}

@article{Chen:2022jhx,
    title         = {{On Galilean conformal bootstrap. Part II. {\ensuremath{\xi}} = 0 sector}},
    author        = {Chen, Bin and Hao, Peng-xiang and Liu, Reiko and Yu, Zhe-fei},
    eprint        = {2207.01474},
    year          = {2022},
    archiveprefix = {arXiv},
    doi           = {10.1007/JHEP12(2022)019},
    journal       = {JHEP},
    pages         = {019},
    primaryclass  = {hep-th},
    volume        = {12},
}

@article{Chang:2022jut,
    title         = {{Shadow celestial amplitudes}},
    author        = {Chang, Chi-Ming and Cui, Wei and Ma, Wen-Jie and Shu, Hongfei and Zou, Hao},
    eprint        = {2210.04725},
    year          = {2023},
    archiveprefix = {arXiv},
    doi           = {10.1007/JHEP02(2023)017},
    journal       = {JHEP},
    pages         = {017},
    primaryclass  = {hep-th},
    volume        = {02},
}

@article{Chang:2022seh,
    title         = {{Missing corner in the sky: massless three-point celestial amplitudes}},
    author        = {Chang, Chi-Ming and Ma, Wen-Jie},
    eprint        = {2212.07025},
    year          = {2023},
    archiveprefix = {arXiv},
    doi           = {10.1007/JHEP04(2023)051},
    journal       = {JHEP},
    pages         = {051},
    primaryclass  = {hep-th},
    volume        = {04},
}

@article{Chen:2023pqf,
    title         = {{Constructing Carrollian field theories from null reduction}},
    author        = {Chen, Bin and Liu, Reiko and Sun, Haowei and Zheng, Yu-fan},
    eprint        = {2301.06011},
    year          = {2023},
    archiveprefix = {arXiv},
    doi           = {10.1007/JHEP11(2023)170},
    journal       = {JHEP},
    pages         = {170},
    primaryclass  = {hep-th},
    volume        = {11},
}

@article{Chang:2023ttm,
    title         = {{Split representation in celestial holography}},
    author        = {Chang, Chi-Ming and Liu, Reiko and Ma, Wen-Jie},
    eprint        = {2311.08736},
    year          = {2023},
    archiveprefix = {arXiv},
    journal       = {},
    month         = {11},
    primaryclass  = {hep-th},
}

@article{Chen:2023naw,
    title         = {{Bulk reconstruction in flat holography}},
    author        = {Chen, Bin and Hu, Zezhou},
    eprint        = {2312.13574},
    year          = {2024},
    archiveprefix = {arXiv},
    doi           = {10.1007/JHEP03(2024)064},
    journal       = {JHEP},
    pages         = {064},
    primaryclass  = {hep-th},
    volume        = {03},
}

@article{Mitra:2024ugt,
    title         = {{Celestial Conformal Primaries in Effective Field Theories}},
    author        = {Mitra, Prahar},
    eprint        = {2402.09256},
    year          = {2024},
    archiveprefix = {arXiv},
    journal       = {},
    month         = {2},
    primaryclass  = {hep-th},
}

@article{Liu:2024lbs,
    title         = {{Massive celestial amplitudes and celestial amplitudes beyond four points}},
    author        = {Liu, Reiko and Ma, Wen-Jie},
    eprint        = {2404.01920},
    year          = {2025},
    archiveprefix = {arXiv},
    doi           = {10.1007/JHEP01(2025)180},
    journal       = {JHEP},
    pages         = {180},
    primaryclass  = {hep-th},
    volume        = {01},
}

@article{Liu:2024vmx,
    title         = {{Celestial optical theorem}},
    author        = {Liu, Reiko and Ma, Wen-Jie},
    eprint        = {2404.18898},
    year          = {2025},
    archiveprefix = {arXiv},
    doi           = {10.1103/PhysRevD.111.025017},
    journal       = {Phys. Rev. D},
    number        = {2},
    pages         = {025017},
    primaryclass  = {hep-th},
    volume        = {111},
}

@article{Chen:2024voz,
    title         = {{Quantization of Carrollian conformal scalar theories}},
    author        = {Chen, Bin and Sun, Haowei and Zheng, Yu-fan},
    eprint        = {2406.17451},
    year          = {2024},
    archiveprefix = {arXiv},
    doi           = {10.1103/PhysRevD.110.125010},
    journal       = {Phys. Rev. D},
    number        = {12},
    pages         = {125010},
    primaryclass  = {hep-th},
    volume        = {110},
}

@article{Liu:2025dhh,
    title         = {{Amplitude from crossing-symmetric celestial OPE}},
    author        = {Liu, Reiko and Ma, Wen-Jie},
    eprint        = {2503.21512},
    year          = {2025},
    archiveprefix = {arXiv},
    journal       = {},
    month         = {3},
    primaryclass  = {hep-th},
}

@article{Nguyen:2025zhg,
    title         = {{Lectures on Carrollian Holography}},
    author        = {Nguyen, Kevin},
    eprint        = {2511.10162},
    year          = {2025},
    archiveprefix = {arXiv},
    journal       = {},
    month         = {11},
    primaryclass  = {hep-th},
}

@article{Liu:2025voe,
    title         = {{Regular celestial amplitudes}},
    author        = {Liu, Reiko and Ma, Wen-Jie},
    eprint        = {2512.05882},
    year          = {2025},
    archiveprefix = {arXiv},
    journal       = {},
    month         = {12},
    primaryclass  = {hep-th},
}

@article{Liu:2026ocv,
    title         = {{Unifying soft and hard dynamics: The hard current algebra in celestial holography}},
    author        = {Liu, Reiko and Ma, Wen-Jie},
    eprint        = {2601.10601},
    year          = {2026},
    archiveprefix = {arXiv},
    journal       = {},
    month         = {1},
    primaryclass  = {hep-th},
}

@book{Morimoto1,
    title     = {An Introduction to Sato's Hyperfunctions},
    author    = {Morimoto, M.},
    year      = {1993},
    isbn      = {9780821887677},
    publisher = {American Mathematical Society},
    series    = {Graduate Studies in Mathematics},
    url       = {https://books.google.com/books?id=pcSumZ4aPX0C},
}

@book{Gelfand2,
    title     = {Generalized Functions, Volume 2},
    author    = {Gel'Fand, I.M. and Shilov, G.E.},
    year      = {2013},
    isbn      = {9781483262307},
    publisher = {Elsevier Science},
    url       = {https://books.google.com/books?id=iMo3BQAAQBAJ},
}

@book{Gelfand1,
    title     = {Generalized Functions, Volume 1},
    author    = {Gel'fand, I.M. and Shilov, G.E.},
    year      = {2014},
    isbn      = {9781483261591},
    publisher = {Elsevier Science},
    url       = {https://books.google.com/books?id=dgnjBQAAQBAJ},
}

@book{Strominger:2017zoo,
    title         = {{Lectures on the Infrared Structure of Gravity and Gauge Theory}},
    author        = {Strominger, Andrew},
    eprint        = {1703.05448},
    year          = {2018},
    archiveprefix = {arXiv},
    isbn          = {978-0-691-17973-5},
    primaryclass  = {hep-th},
    publisher     = {Princeton University Press},
}

@inproceedings{Pasterski:2021raf,
    title         = {{Celestial Holography}},
    author        = {Pasterski, Sabrina and Pate, Monica and Raclariu, Ana-Maria},
    eprint        = {2111.11392},
    year          = {2021},
    archiveprefix = {arXiv},
    booktitle     = {{Snowmass 2021}},
    month         = {11},
    primaryclass  = {hep-th},
}

@misc{kobayashi2015,
    title         = {Branching laws for Verma modules and applications in parabolic geometry. I},
    author        = {Toshiyuki Kobayashi and Bent Ørsted and Petr Somberg and Vladimir Soucek},
    eprint        = {1305.6040},
    year          = {2015},
    archiveprefix = {arXiv},
    primaryclass  = {math.RT},
    url           = {https://arxiv.org/abs/1305.6040},
}

@misc{baker2025,
    title         = {Monitoring Reasoning Models for Misbehavior and the Risks of Promoting Obfuscation},
    author        = {Bowen Baker and Joost Huizinga and Leo Gao and Zehao Dou and Melody Y. Guan and Aleksander Madry and Wojciech Zaremba and Jakub Pachocki and David Farhi},
    eprint        = {2503.11926},
    year          = {2025},
    archiveprefix = {arXiv},
    primaryclass  = {cs.AI},
    url           = {https://arxiv.org/abs/2503.11926},
}

@misc{kalai2025,
    title         = {Why Language Models Hallucinate},
    author        = {Adam Tauman Kalai and Ofir Nachum and Santosh S. Vempala and Edwin Zhang},
    eprint        = {2509.04664},
    year          = {2025},
    archiveprefix = {arXiv},
    primaryclass  = {cs.CL},
    url           = {https://arxiv.org/abs/2509.04664},
}

@misc{cap2000,
    title         = {Bernstein-Gelfand-Gelfand sequences},
    author        = {Andreas Cap and Jan Slovak and Vladimir Soucek},
    eprint        = {math/0001164},
    year          = {2000},
    archiveprefix = {arXiv},
    primaryclass  = {math.DG},
    url           = {https://arxiv.org/abs/math/0001164},
}

\end{document}